\documentclass[epj]{webofc}
\usepackage[varg]{txfonts}   

\usepackage{packages}

\crefname{section}{Sect.}{Sect.}
\Crefname{section}{Section}{Sections}

\graphicspath{%
  {./},%
  {./plots/}
}

\usepackage{macros}

\woctitle{MESON2016 - the 14$^\textrm{th}$ International Workshop on Meson Production, Properties and Interaction}

\begin{document}

\selectlanguage{english}
\title{Meson Spectroscopy at COMPASS}

\author{Boris~Grube\inst{1}\fnsep\thanks{\email{bgrube@tum.de}} \\
  for the COMPASS Collaboration
}

\institute{Institute for Hadronic Structure and Fundamental Symmetries, Physik-Department, Technische Universität München}

\abstract{%
  The goal of the COMPASS experiment at CERN is to study the structure
  and dynamics of hadrons.  The two-stage spectrometer used by the
  experiment has large acceptance and covers a wide kinematic range
  for charged as well as neutral particles and can therefore measure a
  wide range of reactions.  The spectroscopy of light mesons is
  performed with negative (mostly $\pi^-$) and positive ($p$, $\pi^+$)
  hadron beams with a momentum of \SI{190}{\GeVc}.  The light-meson
  spectrum is measured in different final states produced in
  diffractive dissociation reactions with squared four-momentum
  transfer $t$ to the target between \SIlist{0.1;1.0}{\GeVcsq}.  The
  flagship channel is the \threePi final state, for which COMPASS has
  recorded the currently world's largest data sample.  These data not
  only allow to measure the properties of known resonances with high
  precision, but also to observe new states.  Among these is a new
  axial-vector signal, the \PaOne[1420], with unusual properties.
  Novel analysis techniques have been developed to extract also the
  amplitude of the \twoPi subsystem as a function of $3\pi$ mass from
  the data.  The findings are confirmed by the analysis of the
  \threePiN final state.}

\maketitle

\makeatletter
\g@addto@macro\bfseries{\boldmath}
\makeatother

\section{Introduction}
\label{intro}

The COMPASS experiment~\cite{compass} has recorded large data sets of
the diffractive dissociation reaction
$\pi^- + p \to (3\pi)^- + p_\text{recoil}$ using a \SI{190}{\GeVc}
pion beam on a liquid-hydrogen target.  In this process, the beam
hadron is excited to some intermediate three-pion state $X^-$ via
$t$-channel Reggeon exchange with the target.  At \SI{190}{\GeVc} beam
momentum, Pomeron exchange is dominant.  Diffractive reactions are
known to exhibit a rich spectrum of intermediate states $X^-$ and are
a good place to search for states beyond the naive constituent-quark
model.  In the past, several candidates for so-called spin-exotic
mesons, which have \JPC quantum numbers that are forbidden in the
non-relativistic quark model, have been reported in pion-induced
diffraction~\cite{exotic_1,exotic_2}.

The scattering process is characterized by two kinematic variables:
the squared total center-of-mass energy $s$, which is fixed by the
beam energy, and the squared four-momentum transfer to the target
$t = (p_\text{beam} - p_{X})^2 < 0$.  It is customary to use the
reduced four-momentum transfer squared
$\tpr \equiv |t| - |t|_\text{min}$ instead of $t$, where
$|t|_\text{min}$ is the minimum value of $|t|$ for a given invariant
mass of $X^-$.  The analysis is performed in the range
\SIvalRange{0.1}{\tpr}{1.0}{\GeVcsq}.

\begin{figure}[!tb]
  \centering
  \includegraphics[width=0.32\textwidth]{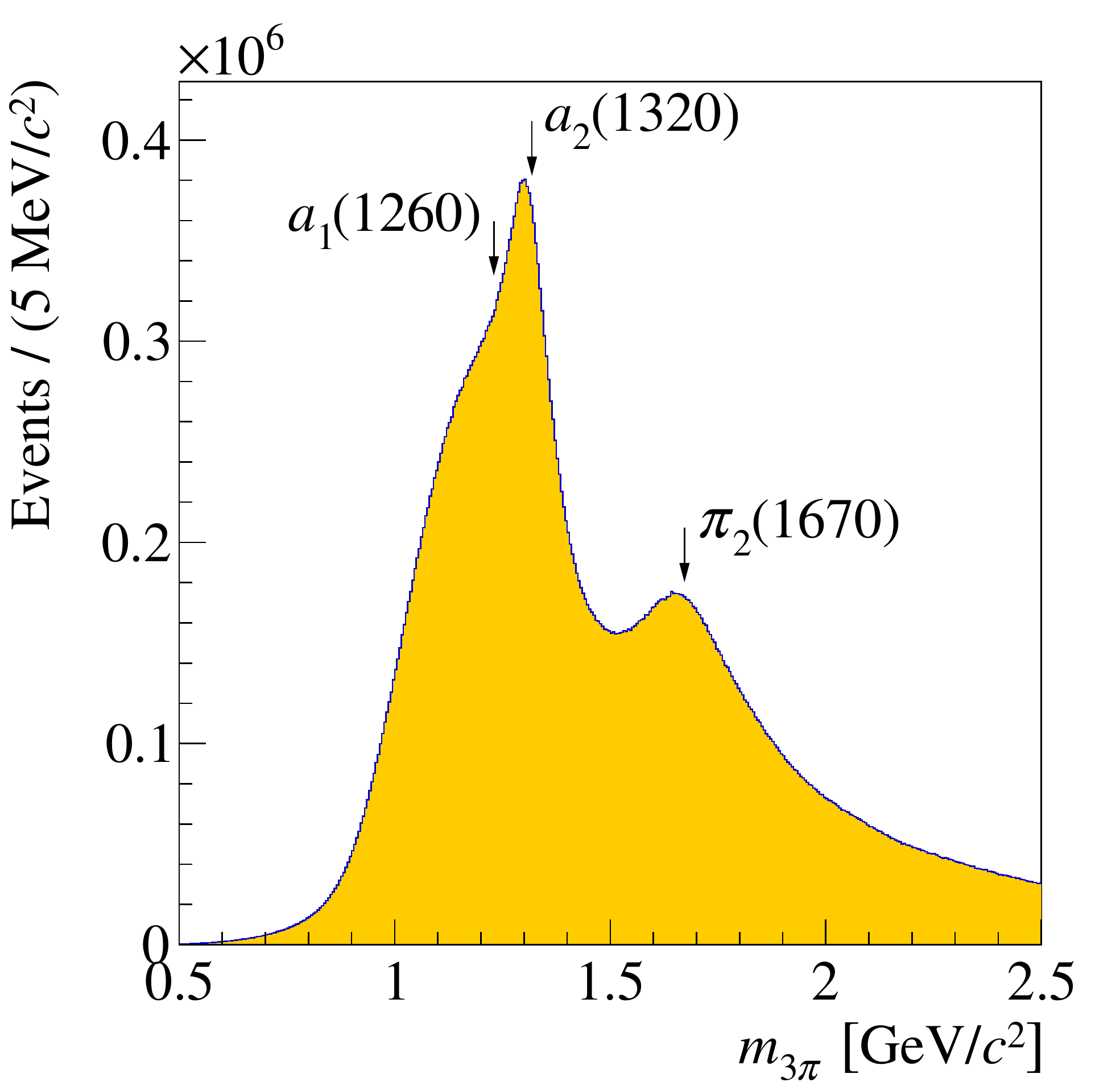}
  \qquad
  \includegraphics[width=0.32\textwidth]{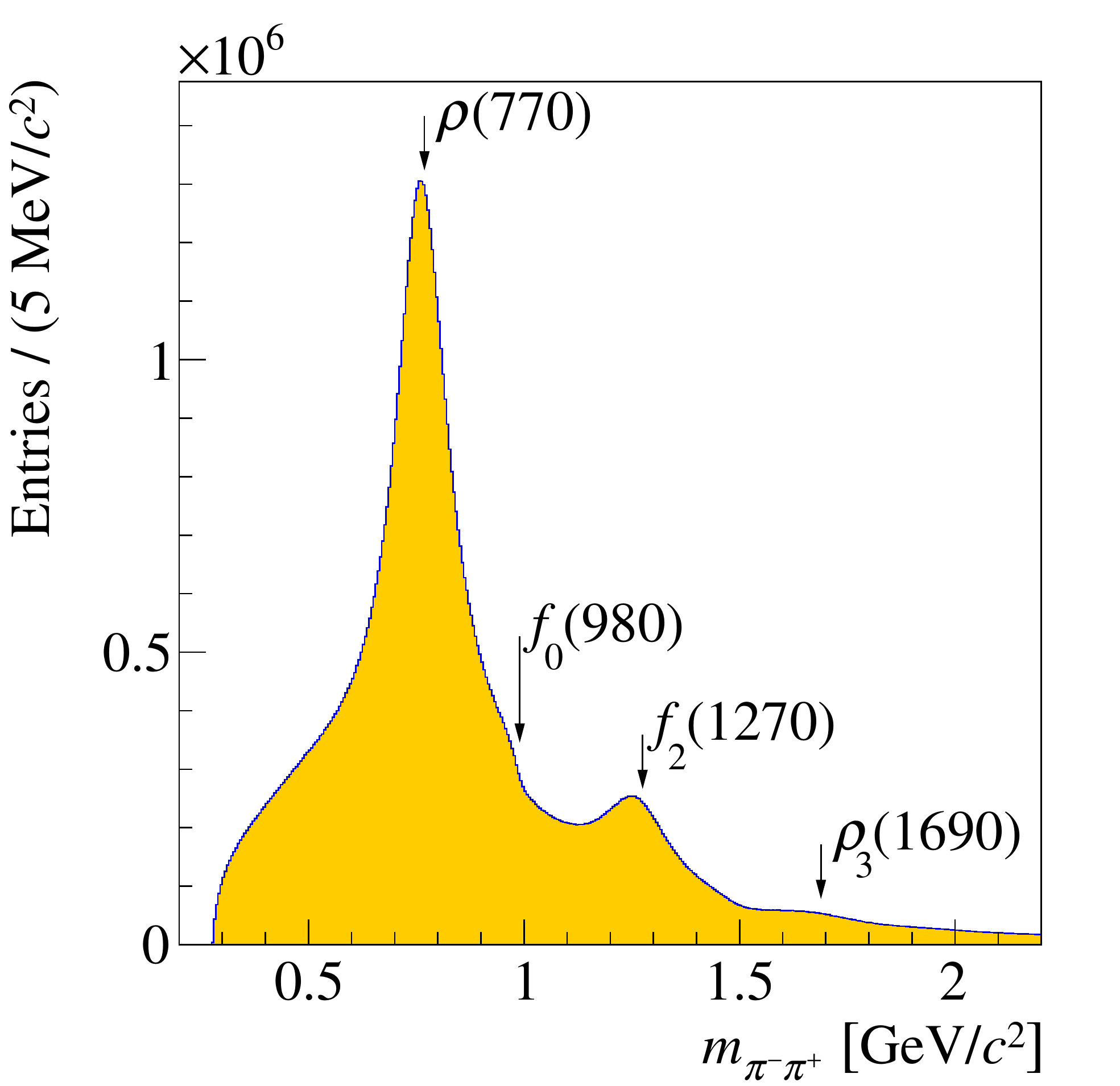}
  \caption{Left: \threePi invariant mass spectrum in the analyzed
    range; Right: invariant mass distribution of the \twoPi subsystem
    (two entries per event). Figures from \refCite{long_paper}.}
  \label{fig:mass}
\end{figure}

In addition to the three final-state pions from the $X^-$ decays, also
the recoiling proton is measured.  This helps to suppress backgrounds
and ensures an exclusive measurement by applying energy and momentum
conservation in the event selection.  After all selection cuts, the
$3\pi$ data samples consist of \num{46e6} \threePi and \num{3.5e6}
\threePiN exclusive events in the analyzed kinematic region of
three-pion mass, \SIvalRange{0.5}{\mThreePi}{2.5}{\GeVcc}.
\Cref{fig:mass} shows the \threePi invariant mass spectrum together
with that of the \twoPi subsystem.  The known pattern of resonances
\PaOne, \PaTwo, and \PpiTwo is seen in the $3\pi$ system along with
\Prho, \PfZero[980], \PfTwo, and \PrhoThree in the \twoPi subsystem.

\section{Partial-Wave Decomposition}

In order to disentangle the different contributing intermediate states
$X^-$, a partial-wave analysis (PWA) is performed.  The PWA of the
$(3\pi)^-$ final states is based on the isobar model, which assumes
that the $X^-$ decays first into an intermediate resonance, which is
called the isobar, and a \enquote{bachelor} pion ($\pi^-$ for the
\threePi final state; $\pi^-$ or $\pi^0$ for \threePiN).  In a second
step, the isobar decays into two pions.  In accordance with the \twoPi
invariant mass spectrum shown in \cref{fig:mass} right and with
analyses by previous experiments, we include \pipiS, \Prho,
\PfZero[980], \PfTwo, \PfZero[1500], and \PrhoThree as isobars into
the fit model.  Here, \pipiS represents the broad component of the
\pipiSW.  Based on the six isobars, we have constructed a set of
partial waves that consists of 88~waves in total, including one
non-interfering isotropic wave representing three uncorrelated pions.
This constitues the largest wave set ever used in an analysis of the
$3\pi$ final state.  The partial-wave decomposition is performed in
narrow bins of the $3\pi$ invariant mass.  Since the data show a
complicated correlation of the \mThreePi and \tpr spectra, each
\mThreePi bin is further subdivided into non-equidistant bins in the
four-momentum transfer \tpr.  For the \threePi channel 11~bins are
used, for the \threePiN final state 8~bins.  With this additional
binning in \tpr, the dependence of the individual partial-wave
amplitudes on the four-momentum transfer can be studied in detail.
The details of the analysis model are described in
\refCite{long_paper}.

The partial-wave amplitudes are extracted from the data as a function
of \mThreePi and \tpr by fitting the five-dimensional kinematic
distributions of the outgoing three pions.  The amplitudes do contain
information not only about the partial-wave intensities, but also
about the relative phases of the partial waves.  The latter are
crucial for resonance extraction.  The three-pion partial waves are
defined by the quantum numbers of the $X^-$ (spin~$J$, parity~$P$,
$C$-parity, absolute value~$M$ of the spin projection), the naturality
$\refl = \pm 1$ of the exchange particle, the isobar, and the orbital
angular momentum~$L$ between the isobar and the bachelor pion.  These
quantities are summarized in the partial-wave notation
\wave{J}{PC}{M}{\refl}{\text{[isobar]}}{L}.  Since at the used beam
energies Pomeron exchange is dominant, 80~of the 88~partial waves in
the model have $\refl = +1$.

\subsection{The \PaOne[1420]}

\begin{figure}[!tb]
  \centering
  \includegraphics[width=0.32\textwidth]{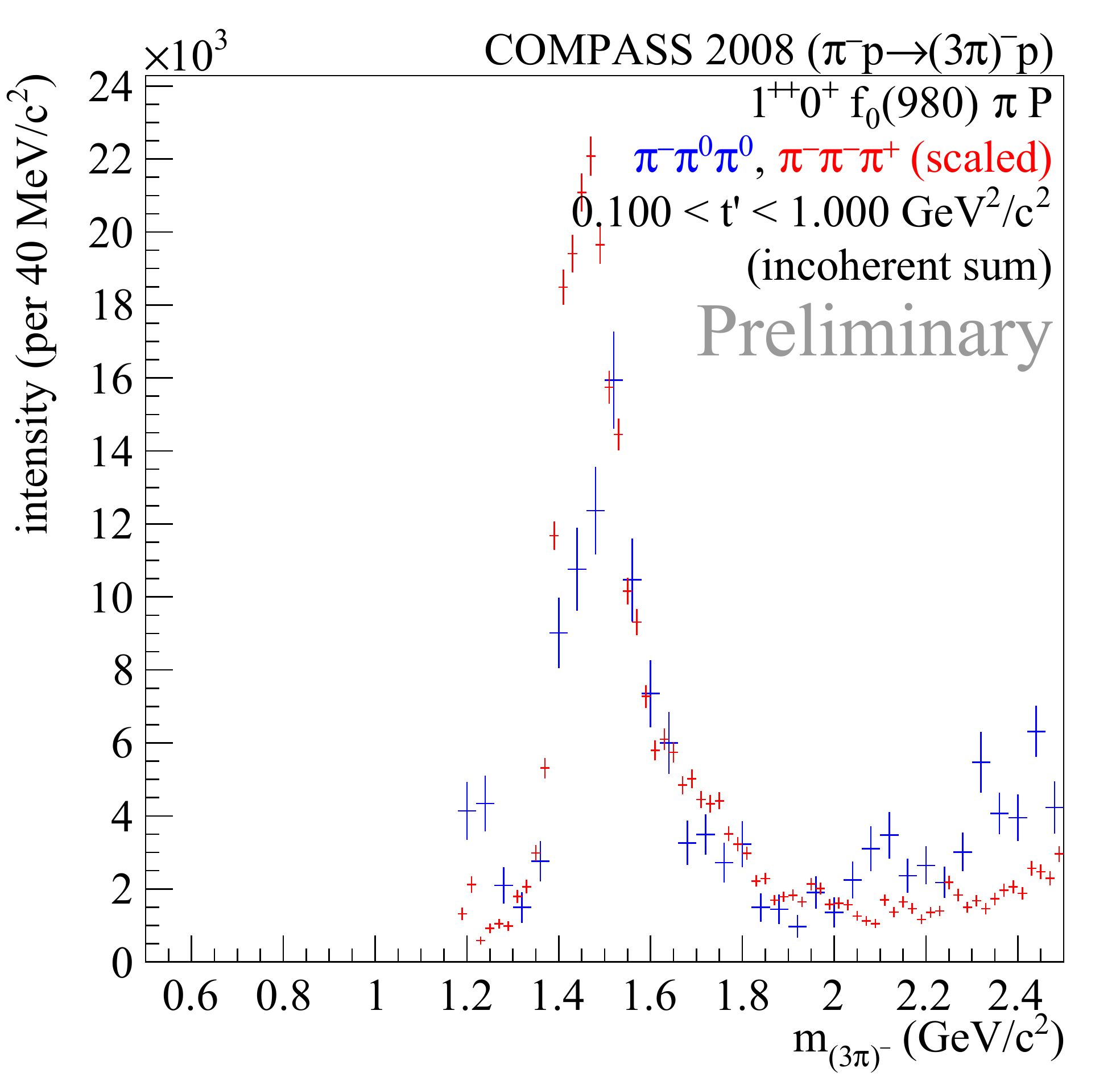}
  \qquad
  \includegraphics[width=0.32\textwidth]{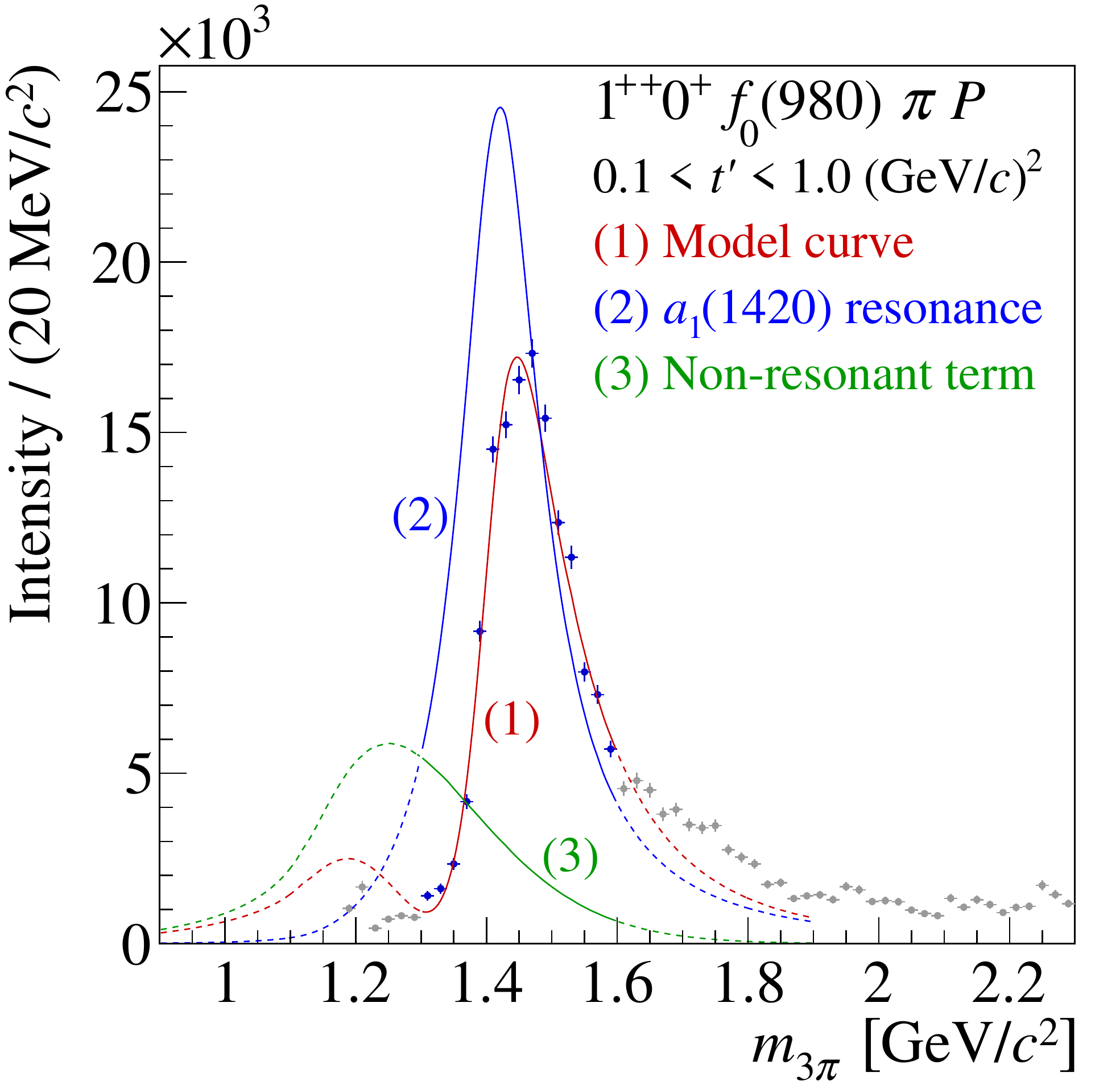}
  \caption{Left: Intensity of the \wave{1}{++}{0}{+}{\PfZero[980]}{P}
    wave summed over all \tpr bins for the \threePiN (blue) and the
    \threePi (red, scaled to the intensity integral of the \threePiN
    channel) final states.  Right: Result of a resonance-model fit to
    the \threePi data~\cite{a1_1420}.  The data points correspond to
    the red points in the left figure.}
  \label{fig:1pp_f0_intens}
\end{figure}

\begin{figure}[!tb]
  \centering
  \includegraphics[width=0.32\textwidth]{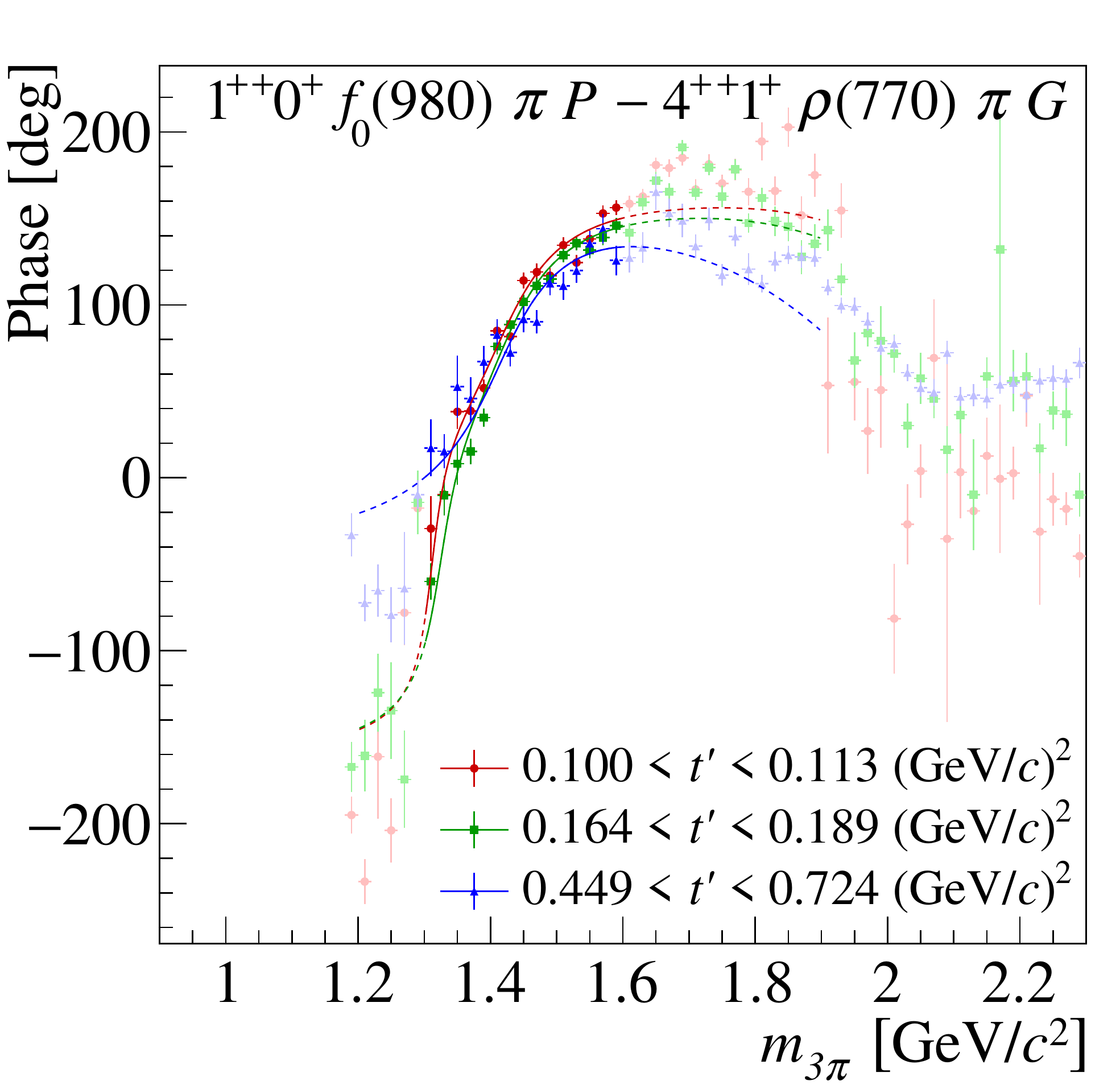}
  \qquad
  \includegraphics[width=0.32\textwidth]{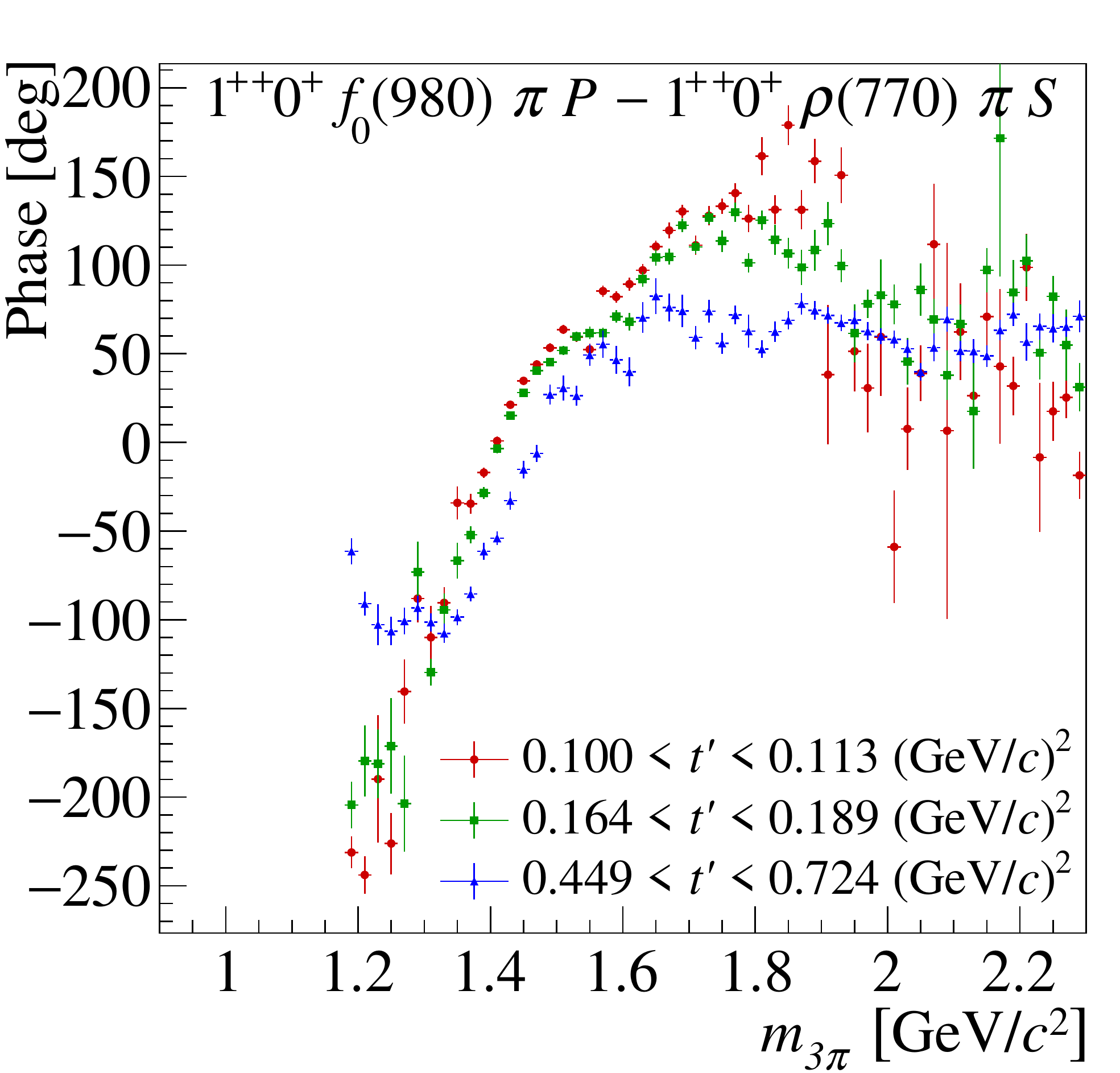}
  \caption{Examples for relative phases of the
    \wave{1}{++}{0}{+}{\PfZero[980]}{P} wave \wrt the
    \wave{4}{++}{1}{+}{\Prho}{G} (left) and the
    \wave{1}{++}{0}{+}{\Prho}{S} wave (right).  The phases are shown
    for three different \tpr regions indicated by the color. Figures
    from \refCite{a1_1420}.}
  \label{fig:1pp_f0_phase}
\end{figure}

A surprising find in the COMPASS data is a pronounced narrow peak at
about \SI{1.4}{\GeVcc} in the \wave{1}{++}{0}{+}{\PfZero[980]}{P} wave
(see \cref{fig:1pp_f0_intens}).  The peak is observed with similar
shape in the \threePi and \threePiN channels and is robust against
variations of the PWA model.  In addition to the peak in the
partial-wave intensity, rapid phase variations \wrt most waves are
observed in the \SI{1.4}{\GeVcc} region (see \cref{fig:1pp_f0_phase}).
The phase motion as well as the peak shape change only little with
\tpr.

In order to test the compatibility of the signal with a Breit-Wigner
resonance, a resonance-model fit was performed using a novel method,
where the intensities and relative phases of three waves
[\wave{1}{++}{0}{+}{\PfZero[980]}{P}, \wave{2}{++}{1}{+}{\Prho}{D},
and \wave{4}{++}{1}{+}{\Prho}{G}] were fit simultaneously in all
11~\tpr bins~\cite{a1_1420}.  Forcing the resonance parameters to be
the same across all \tpr bins leads to an improved separation of
resonant and non-resonant contribution as compared to previous
analyses that did not incorporate the \tpr information.  The
Breit-Wigner model describes the peak in the
\wave{1}{++}{0}{+}{\PfZero[980]}{P} wave well and yields a mass of
$m_0 = \SIaerr{1414}{15}{13}{\MeVcc}$ and a width of
$\Gamma_0 = \SIaerr{153}{8}{23}{\MeVcc}$ for the \PaOne[1420].  Due to
the high statistical precision of the data, the uncertainties are
dominated by systematic effects.

The \PaOne[1420] signal is remarkable in many ways.  It appears in a
mass region that is well studied since decades.  However, previous
experiments were unable to see the peak, because it contributes only
\SI{0.25}{\percent} to the total intensity.  The \PaOne[1420] is very
close in mass to the $1^{++}$ ground state, the \PaOne.  But it has a
much smaller width than the \PaOne.  The \PaOne[1420] peak is seen
only in the $\PfZero[980]\,\pi$ decay mode of the $1^{++}$ waves and lies
suspiciously close to the $K\, \PKbar^*(892)$ threshold.

The nature of the \PaOne[1420] is still unclear and several
interpretations were proposed.  It could be the isospin partner to the
$f_1(1420)$.  It was also described as a two-quark-tetraquark mixed
state~\cite{wang} and a tetraquark with mixed flavor
symmetry~\cite{chen}.  Other models do not require an additional
resonance: the authors of \refsCite{berger1,berger2} propose resonant
re-scattering corrections in the Deck process as an explanation,
whereas \refCite{bonn} suggests a branching point in the triangular
rescattering diagram for
$\PaOne \to K\, \PKbar^*(892) \to K\, \PKbar\, \pi \to \PfZero[980]\,
\pi$.
The results of the latter calculation were confirmed by the authors of
\refCite{oset}.  Triangle singularities were also proposed as an
explanation for the narrow \Peta[1405]~\cite{wu1,wu2}, for some of the
near-threshold $XYZ$ heavy-quark states (see \eg\
\refCite{triangle_xyz}), and for the pentaquark candidate $P_c(4450)$
recently found by LHCb~\cite{lhcb_pentaquark,triangle_pentaquark}.
More detailed studies are needed in order to distinguish between the
different models for the \PaOne[1420].

\subsection{Extraction of \pipiSW Isobar Amplitudes from Data}

The PWA of the $3\pi$ system is based on the isobar model, where fixed
amplitudes are used for the description of the \twoPi intermediate
states.  However, we cannot exclude that the fit results are biased by
the employed isobar parametrizations.  This is true in particular for
the isoscalar $\JPC = 0^{++}$ isobars.  In the PWA model, a broad
\pipiSW component is used, the parametrization of which is extracted
from \pipiSW elastic-scattering data~\cite{amp}.  In addition, the
\PfZero[980], described by a Flatt\'e form~\cite{flatte}, and the
\PfZero[1500], parametrized by a relativistic Breit-Wigner amplitude,
are included as isobars.  In order to study possible bias due to these
parametrizations and to ensure that the observed \PaOne[1420] signal
is truly related to the narrow \PfZero[980], a novel analysis method
inspired by \refCite{e791} was developed~\cite{long_paper}.  In this
so-called \emph{freed-isobar} analysis, the three fixed
parametrizations for the $0^{++}$ isobar amplitudes are replaced by a
set of piecewise constant complex-valued functions that fully cover
the allowed two-pion mass range.  This way the whole $0^{++}$ isobar
amplitude is extracted as a function of the $3\pi$ mass.  In contrast
to the conventional isobar approach, which uses the same isobar
parametrization in different partial waves, the freed-isobar method
permits different isobar amplitudes for different intermediate states
$X^-$.  A more detailed description of the analysis method can be
found in \refCite{long_paper}.

\begin{figure}[!tb]
  \centering
  \includegraphics[width=0.32\textwidth]{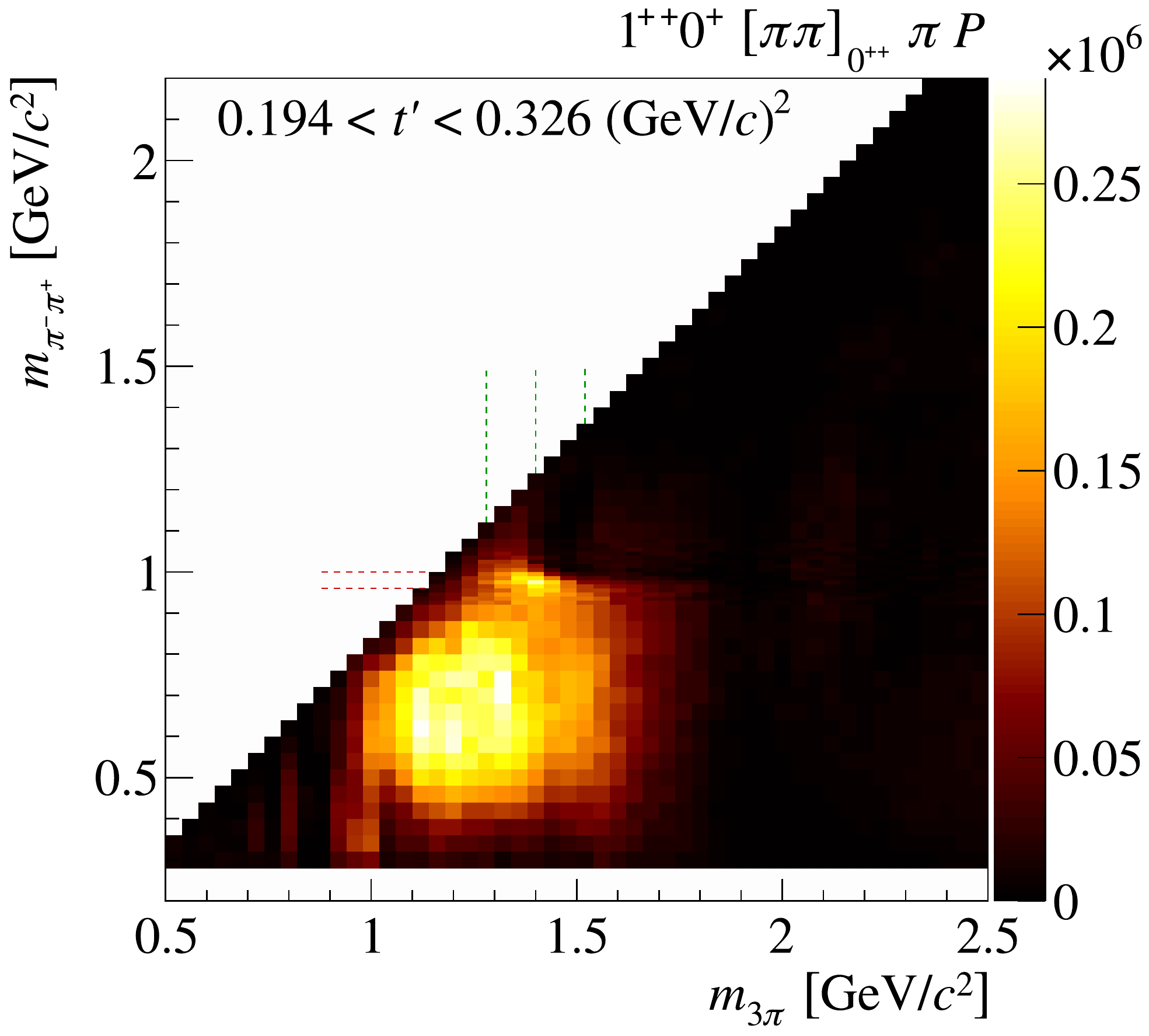}
  \includegraphics[width=0.32\textwidth]{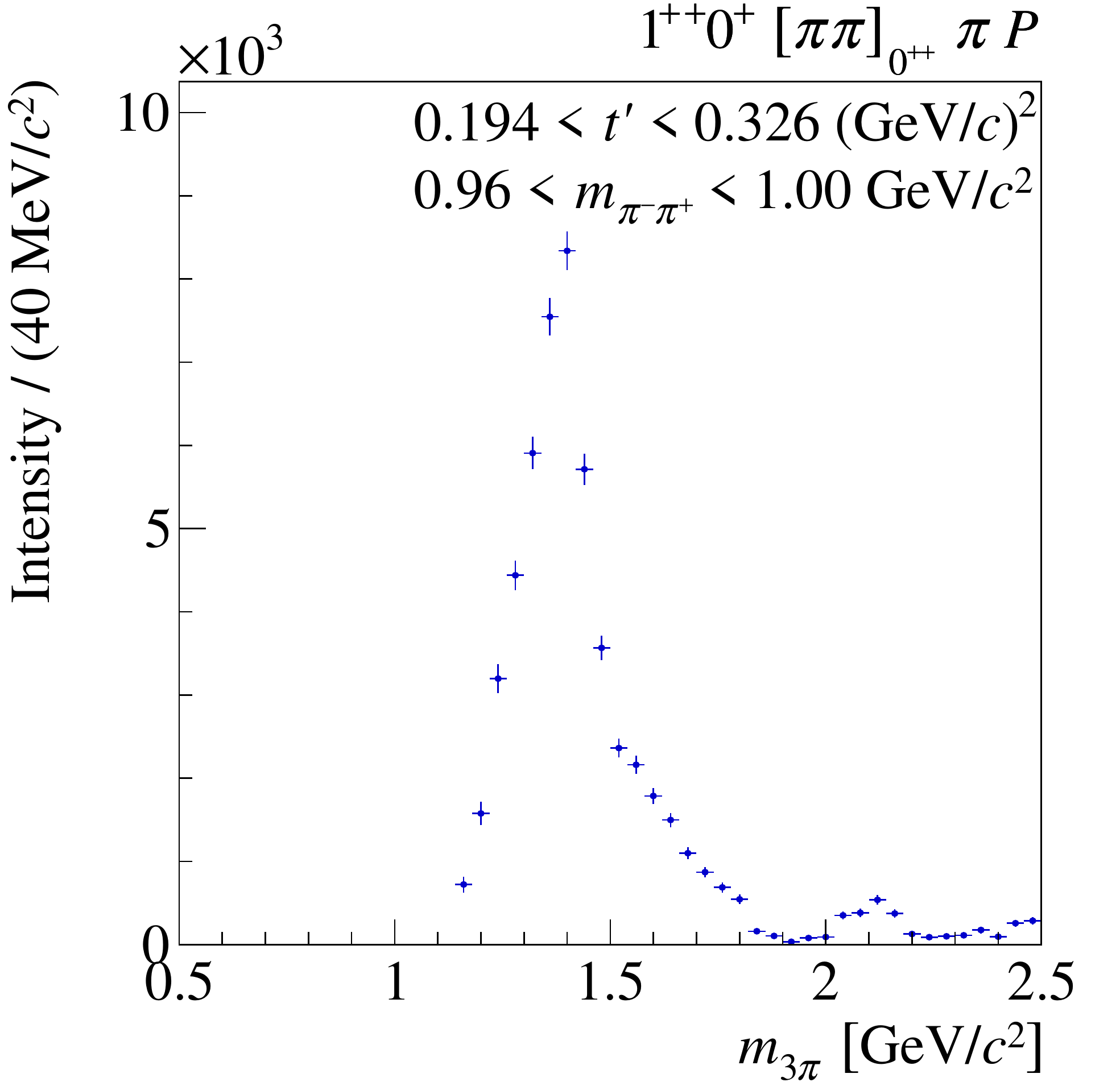}
  \includegraphics[width=0.32\textwidth]{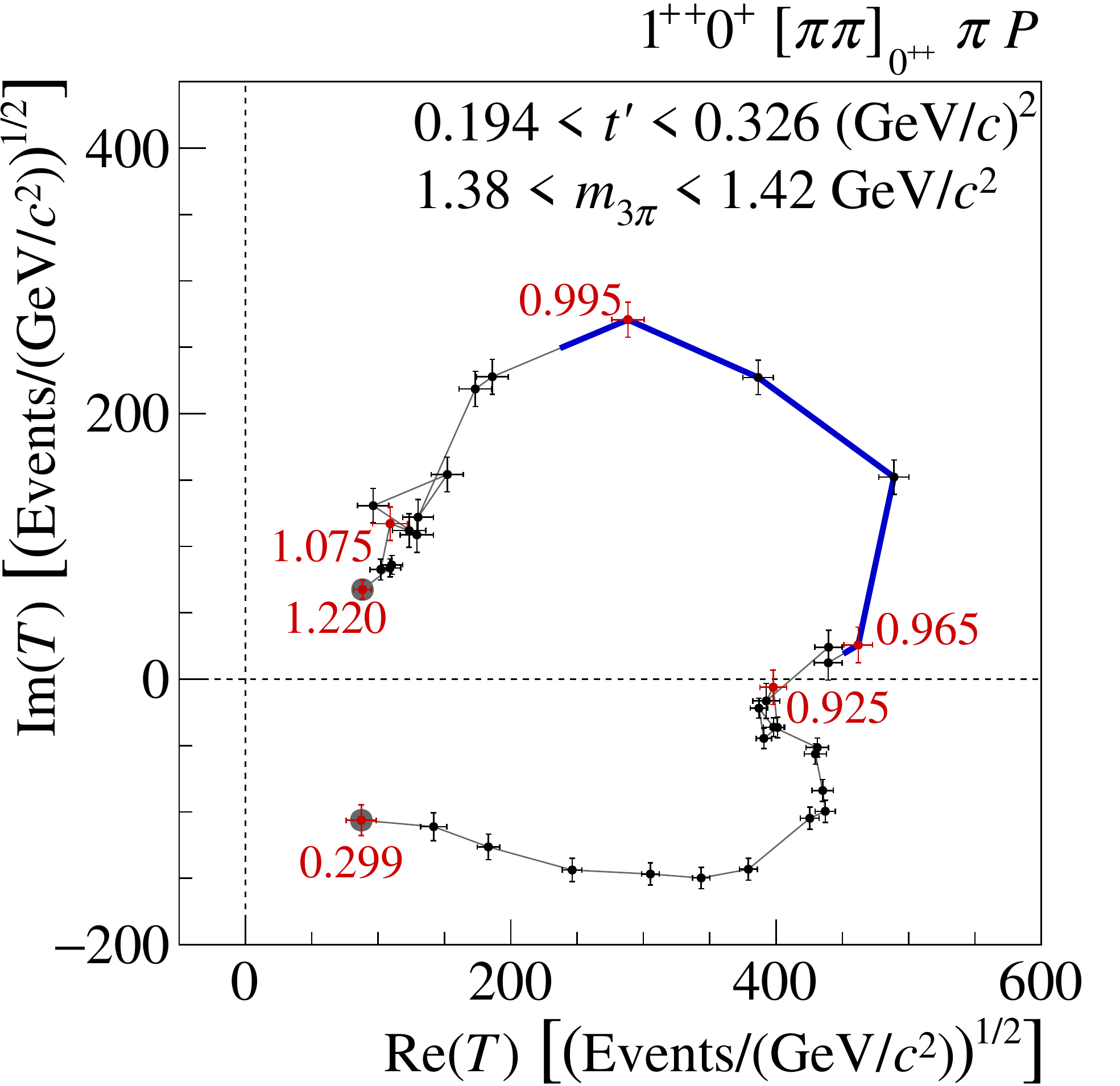}
  \caption{Left: Intensity of the \pipiSW component of the
    $\JPCMrefl = 1^{++}\,0^+$ partial wave resulting from the
    freed-isobar fit as a function of \mTwoPi and \mThreePi.  Center:
    intensity as a function of \mThreePi summed over the \mTwoPi
    interval around the \PfZero[980] indicated by the pair of
    horizontal dashed lines in the top figure.  Right: Argand diagram
    representing the \mTwoPi dependence of the partial-wave amplitude
    for the $3\pi$ mass bin at the \PaOne[1420] measured \wrt the
    \wave{1}{++}{0}{+}{\Prho}{S} wave.  Figures from
    \refCite{long_paper}.}
  \label{fig:freed-isobar}
\end{figure}

The freed-isobar method leads to a reduced model bias and gives
additional information about the \twoPi subsystem at the cost of a
considerable increase in the number of free parameters in the PWA fit.
Thus, even for large data sets, the freed-isobar approach can only be
applied to a subset of partial waves.  We performed a freed-isobar
PWA, where the fixed parametrizations of the broad \pipiSW component,
of the \PfZero[980], and of the \PfZero[1500] were replaced by
piece-wise constant isobar amplitudes for the $3\pi$ partial waves
\wave{0}{-+}{0}{+}{\pipiSF}{S}, \wave{1}{++}{0}{+}{\pipiSF}{P}, and
\wave{2}{-+}{0}{+}{\pipiSF}{D}.  \Cref{fig:freed-isobar} left shows
the two-dimensional intensity distribution of the
\wave{1}{++}{0}{+}{\pipiSF}{P} wave as a function of \mTwoPi and
\mThreePi.  The distribution exhibits a broad maximum around
$\mThreePi = \SI{1.2}{\GeVcc}$ and between \SIlist{0.6;0.8}{\GeVcc} in
\mTwoPi, which shows a pronounced \tpr dependence and therefore is
probably mainly of non-resonant origin.  A smaller peak is observed in
the \PfZero[980] region at $\mThreePi \approx \SI{1.4}{\GeVcc}$.  This
peak is more obvious in \cref{fig:freed-isobar} center, which shows
the intensity distribution summed over the two-pion mass interval
around the \PfZero[980] as indicated by the pair of horizontal dashed
lines in \cref{fig:freed-isobar} left.  The peak is similar in
position and shape to the \PaOne[1420] peak in the
\wave{1}{++}{0}{+}{\PfZero[980]}{P} wave
(cf. \cref{fig:1pp_f0_intens}).  The resonant nature of the
\PfZero[980] becomes apparent in \cref{fig:freed-isobar} right, which
shows the \mTwoPi dependence of the extracted amplitude at the
\PaOne[1420] peak in form of an Argand diagram.  The phase is measured
\wrt the \wave{1}{++}{0}{+}{\Prho}{S} wave.  The \PfZero[980]
contribution shows up as a semicircle-like structure (highlighted by
the blue line) with a shifted origin.  This demonstrates that the
observed \PaOne[1420] signal in the $\PfZero[980]\,\pi$ decay mode is
not an artifact of the $0^{++}$ isobar parametrizations used in the
conventional PWA method.  More results of the freed-isobar PWA are
discussed in \refCite{long_paper}.

\subsection{The $\JPC = 1^{-+}$ Spin-Exotic Wave}

\begin{figure}[!tb]
  \centering
  \includegraphics[width=0.32\textwidth]{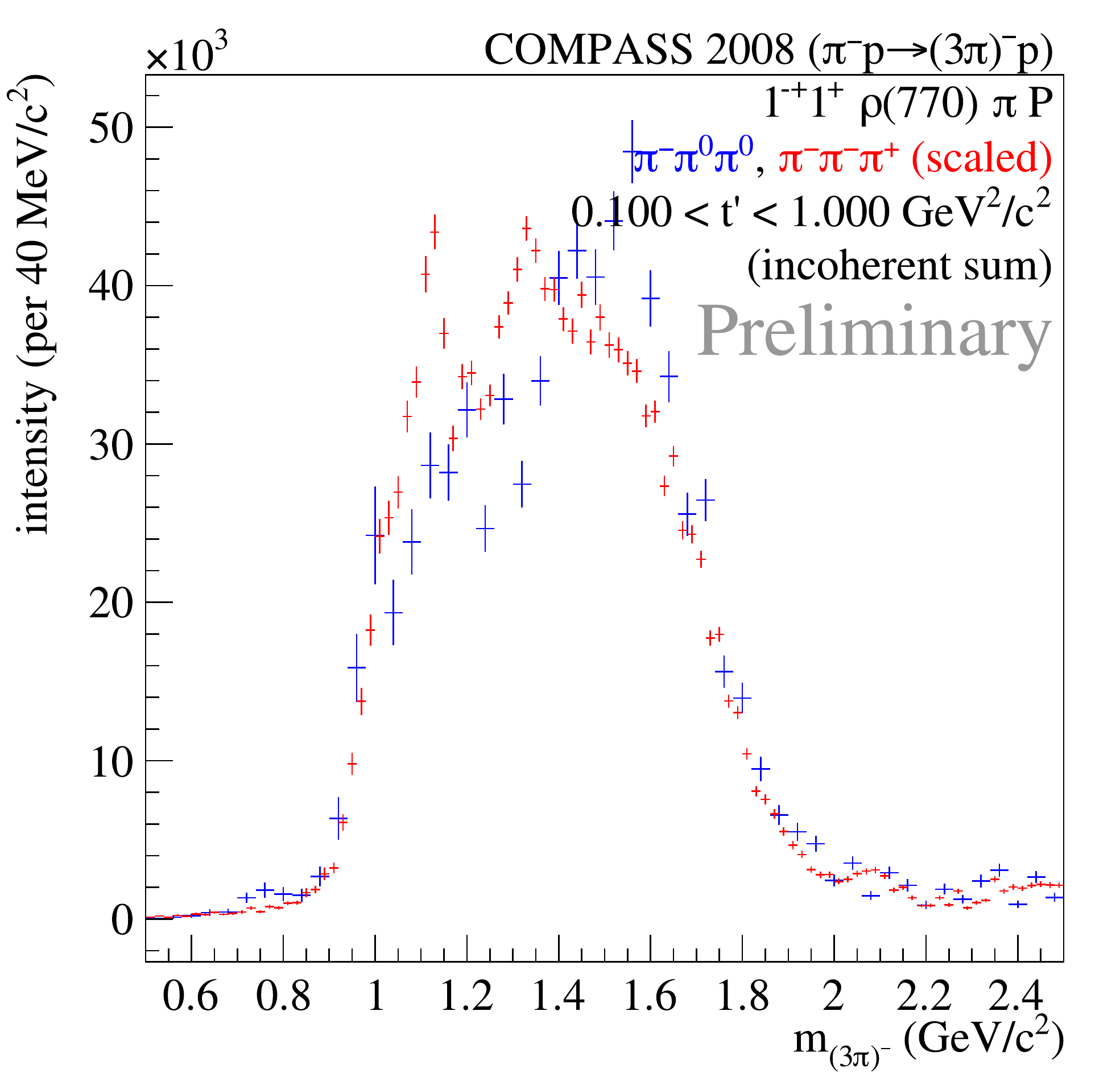}
  \includegraphics[width=0.32\textwidth]{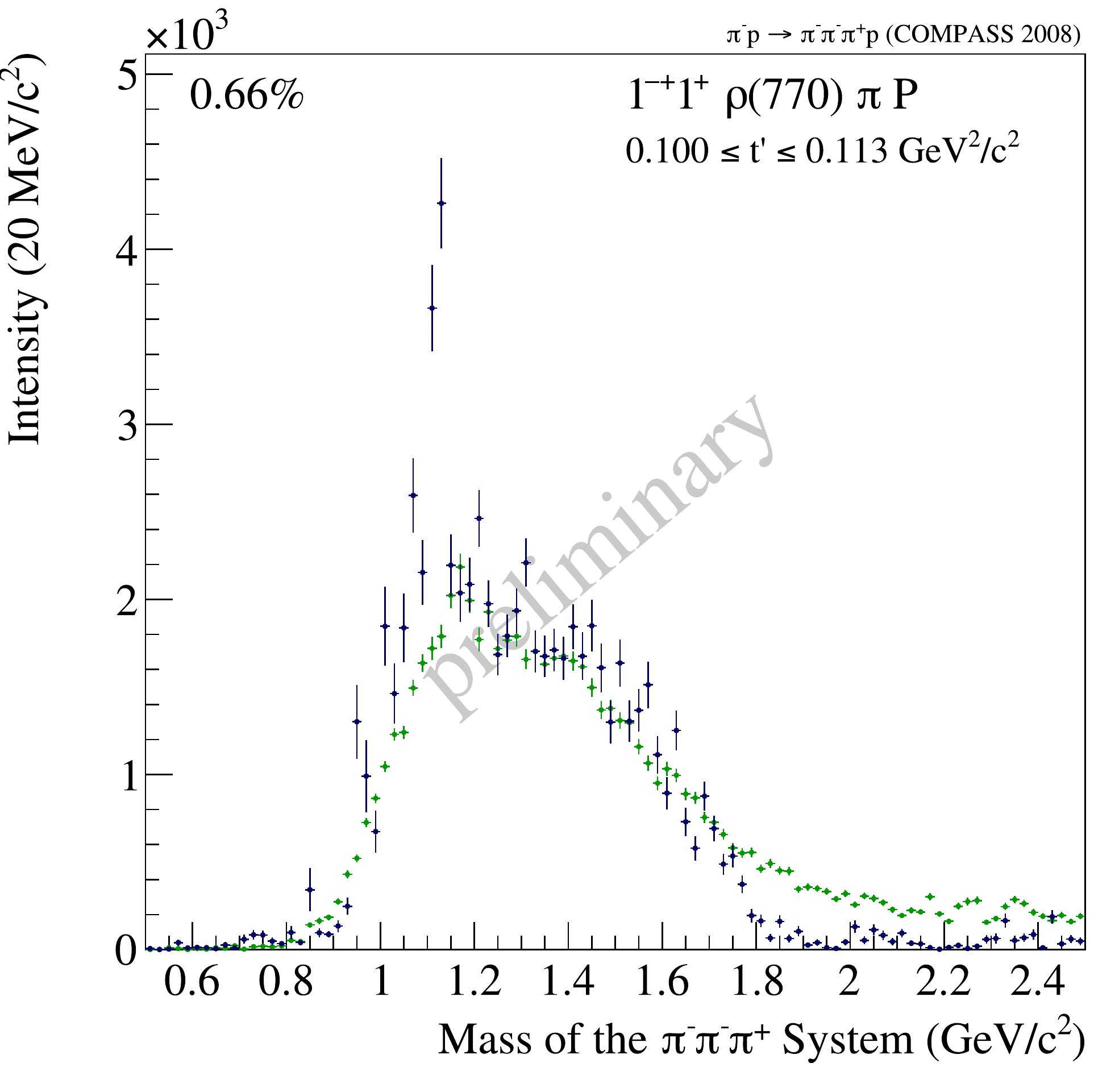}
  \includegraphics[width=0.32\textwidth]{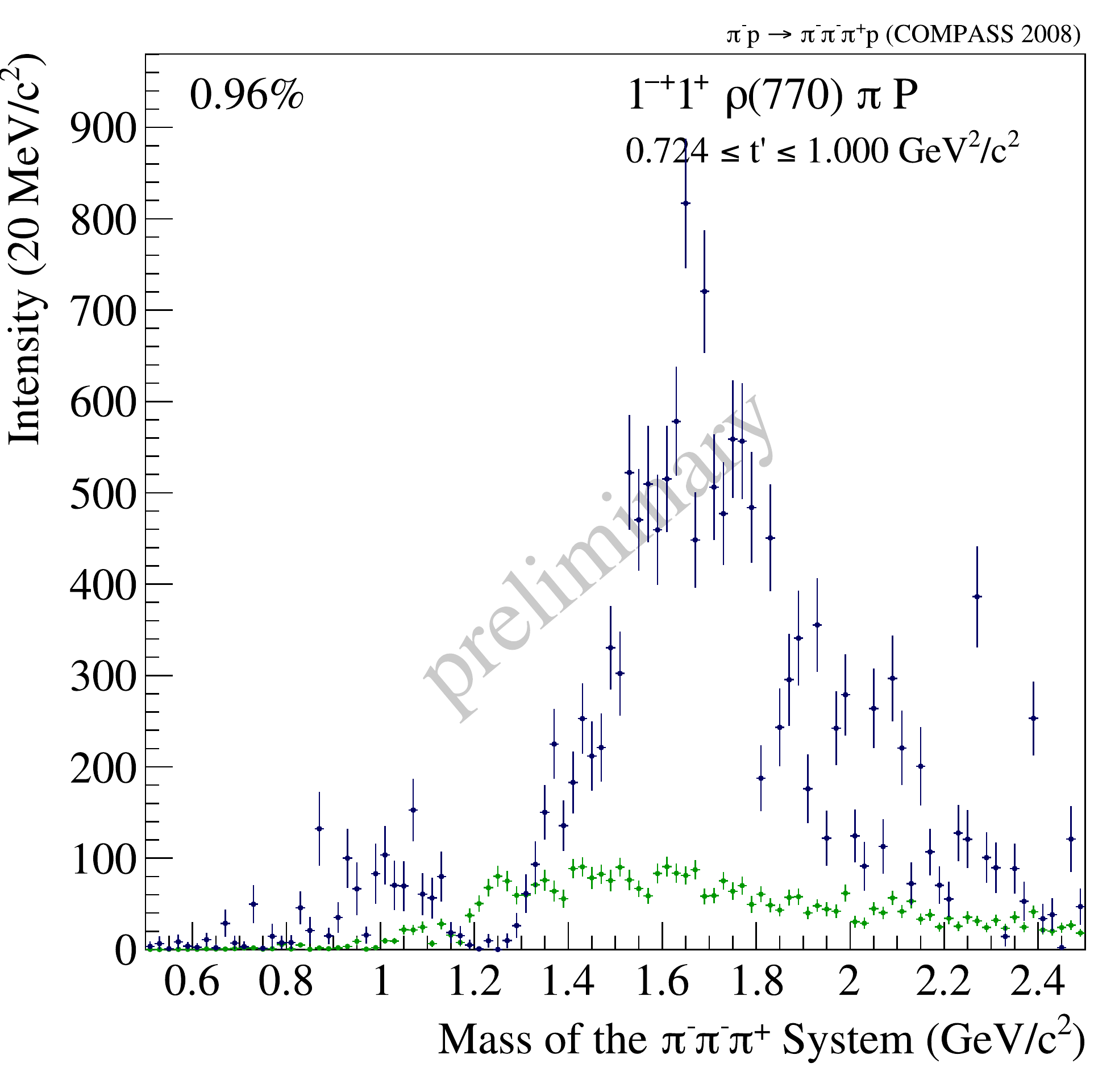}
  \caption{Intensity of the \wave{1}{-+}{1}{+}{\Prho}{P} wave.  Left:
    summed over all \tpr bins for the \threePi (red) and the \threePiN
    (blue) final state.  Center and right panels show the intensity
    for the \threePi final state (dark blue) in different regions of
    \tpr (center: low \tpr; right: high \tpr).  The partial-wave
    projections of Monte-Carlo data generated according to a model of
    the Deck effect are overlaid in green.}
  \label{fig:1mp}
\end{figure}

The 88-wave model also contains waves with exotic \JPC quantum
numbers, that are forbidden in the non-relativistic quark model.  The
most interesting of these waves is the \wave{1}{-+}{1}{+}{\Prho}{P}
wave, which contributes less than \SI{1}{\percent} to the total
intensity.  Previous analyses claimed a resonance, the \PpiOne[1600],
at about \SI{1.6}{\GeVcc} in this channel~\cite{bnl_1,compass_pb}.
\Cref{fig:1mp} left shows the intensity sum over all \tpr bins of this
partial wave for the two final states (\threePi in red, \threePiN in
blue).  The two distributions are scaled to have the same integral.
Both decay channels are in fair agreement and exhibit a broad
enhancement extending from about \SIrange{1.0}{1.8}{\GeVcc} in
\mThreePi.  In the \SIrange{1.0}{1.2}{\GeVcc} mass range, the
intensity depends strongly on the details of the fit model.  Peak-like
structures in this region are probably due to cross talk induced by
imperfections of the applied PWA model.

A remarkable change of the shape of the intensity spectrum of the
\wave{1}{-+}{1}{+}{\Prho}{P} wave with \tpr is observed (see dark blue
points in \cref{fig:1mp} center and right).  At values of \tpr below
about \SI{0.3}{\GeVcsq}, we observe no indication of a resonance peak
around $\mThreePi = \SI{1.6}{\GeVcc}$, where we would expect the
\PpiOne[1600].  However, for the \tpr bins in the interval
\SIvalRange{0.449}{\tpr}{1.000}{\GeVcsq}, the observed intensities
exhibit a very different shape as compared to the low-\tpr region,
with a peak structure emerging at about \SI{1.6}{\GeVcc} and the
intensity at lower masses disappearing rapidly with increasing \tpr.
This is in contrast to clean resonance signals like the \PaTwo in the
\wave{2}{++}{1}{+}{\Prho}{D} wave, which, as expected, do not change
their shape with \tpr.  The observed \tpr behavior of the $1^{-+}$
intensity is therefore a strong indication that non-resonant
contributions play a dominant role.

It is believed that the non-resonant contribution in the $1^{-+}$ wave
originates predominantly from the Deck effect, in which the incoming
beam pion dissociates into the isobar and an off-shell pion that
scatters off the target proton to become on-shell~\cite{deck}.  As a
first step towards a better understanding of the non-resonant
contribution, Monte-Carlo data were generated that are distributed
according to a model of the Deck effect.  The model employed here is
very similar to that used in \refCite{accmor_deck}.  The partial-wave
projection of these Monte Carlo data is shown as green points in
\cref{fig:1mp} center and right.  In order to compare the intensities
of real data and the Deck-model pseudo data, the Monte Carlo data are
scaled to the \tpr-summed intensity of the $1^{-+}$ wave as observed
in real data.  At values of \tpr below about \SI{0.3}{\GeVcsq}, the
intensity distributions of real data and Deck Monte Carlo exhibit
strong similarities suggesting that the observed intensity might be
saturated by the Deck effect.  Starting from
$\tpr \approx \SI{0.4}{\GeVcsq}$, the spectral shapes for Deck pseudo
data and real data deviate from each other with the differences
increasing towards larger values of \tpr.  This leaves room for a
potential resonance signal.  It should be noted, however, that the
Deck pseudo data contain no resonant contributions.  Therefore,
potential interference effects between the resonant and non-resonant
amplitudes cannot be assessed in this simple approach.

\begin{acknowledgement}
  This work was supported by the BMBF, the Maier-Leibnitz-Laboratorium
  (MLL), the DFG Cluster of Excellence Exc153 \enquote{Origin and
    Structure of the Universe}, and the computing facilities of the
  Computational Center for Particle and Astrophysics (C2PAP).
\end{acknowledgement}

\bibliography{bgrube}

\begin{thebibliography}{23}

\bibitem{compass}
P.~Abbon et~al. (COMPASS Collaboration), Nucl. Instrum. Methods Phys. Res.,
  Sect. A \textbf{779}, 69 (2015)

\bibitem{exotic_1}
C.~Meyer et~al., Phys. Rev. C \textbf{82}, 025208 (2010)

\bibitem{exotic_2}
E.~Klempt et~al., Phys. Rept. \textbf{454}, 1 (2007)

\bibitem{long_paper}
C.~Adolph et~al. (COMPASS Collaboration), arXiv:1509.00992, \textit{submitted
  to} Phys. Rev. D  (2015)

\bibitem{a1_1420}
C.~Adolph et~al. (COMPASS Collaboration), Phys. Rev. Lett. \textbf{115}, 082001
  (2015)

\bibitem{wang}
Z.G. Wang, arXiv:1401.1134  (2014)

\bibitem{chen}
H.X. Chen et~al., Phys. Rev. D \textbf{91}, 094022 (2015)

\bibitem{berger1}
J.L. Basdevant, E.L. Berger, Phys. Rev. Lett. \textbf{114}, 192001 (2015)

\bibitem{berger2}
J.L. Basdevant, E.L. Berger, arXiv:1501.04643  (2015)

\bibitem{bonn}
M.~Mikhasenko et~al., Phys. Rev. D \textbf{91}, 094015 (2015)

\bibitem{oset}
F.~Aceti, L.R. Dai, E.~Oset, arXiv:1606.06893  (2016)

\bibitem{wu1}
J.J. Wu, X.H. Liu, Q.~Zhao, B.S. Zou, Phys. Rev. Lett. \textbf{108}, 081803
  (2012)

\bibitem{wu2}
X.G. Wu, J.J. Wu, Q.~Zhao, B.S. Zou, Phys. Rev. \textbf{D87}, 014023 (2013)

\bibitem{triangle_xyz}
A.P. Szczepaniak, Phys. Lett. \textbf{B747}, 410 (2015)

\bibitem{lhcb_pentaquark}
R.~Aaij et~al. (LHCb Collaboration), Phys. Rev. Lett. \textbf{115}, 072001
  (2015)

\bibitem{triangle_pentaquark}
F.K. Guo, U.G. Meißner, W.~Wang, Z.~Yang, Phys. Rev. \textbf{D92}, 071502
  (2015)

\bibitem{amp}
K.L. Au, D.~Morgan, M.R. Pennington, Phys. Rev. D \textbf{35}, 1633 (1987)

\bibitem{flatte}
M.~Ablikim et~al. (BES Collaboration), Phys. Lett. B \textbf{607}, 243 (2005)

\bibitem{e791}
E.M. Aitala et~al. (E791 Collaboration), Phys. Rev. D \textbf{73}, 032004
  (2006)

\bibitem{bnl_1}
S.U. Chung et~al. (E852 Collaboration), Phys. Rev. D \textbf{60}, 092001 (1999)

\bibitem{compass_pb}
M.~Alekseev et~al. (COMPASS Collaboration), Phys. Rev. Lett. \textbf{104},
  241803 (2010)

\bibitem{deck}
R.T. Deck, Phys. Rev. Lett. \textbf{13}, 169 (1964)

\bibitem{accmor_deck}
C.~Daum et~al. (ACCMOR Collaboration), Nucl. Phys. B \textbf{182}, 269 (1981)

\end{thebibliography}

\end{document}